# Wind Shear and Turbulence on Titan : Huygens Analysis

Ralph D. Lorenz*

Johns Hopkins University Applied Physics Laboratory, 11100 Johns Hopkins Road,

Laurel, MD 20723, USA

* Corresponding author  tel: +1 443 778 2903   fax: +1 443 778 8939  email: ralph.lorenz@jhuapl.edu

Highlights:

- Wind shear during Huygens descent was within predicted range
- Shear reached terrestrial aviation 'Light Turbulence' levels
- Doppler and probe tilt indicate ~0.2 m/s fluctuations in lowest 4km
- Simple AR(1) model reproduces observed turbulence characteristics

Abstract

Wind shear measured by Doppler tracking of the Huygens probe is evaluated, and found to be within the range anticipated by pre-flight assessments (namely less than two times the Brunt-Vaisala frequency).  The strongest large-scale shear encountered was ~5 m/s/km, a level associated with 'Light' turbulence in terrestrial aviation.   Near-surface winds  (below 4km) have small-scale fluctuations of ~0.2 m/s , indicated both by probe tilt and Doppler tracking, and the characteristics of the fluctuation, of interest for future missions to Titan, can be reproduced with a simple autoregressive (AR(1)) model. The turbulent dissipation rate at an altitude of ~500m is found to be 16 cm$^2$/sec$^3$, which may be a useful benchmark for atmospheric circulation models.

1. Introduction

Wind shear, the spatial gradient of wind speed, is of significant practical interest in planetary exploration and in terrestrial aviation. A vehicle moving relative to the air will encounter time-varying winds which can excite movements which may blur images (Karkoschka, 2016), affect the strength of the radio signal (Dzierma et al., 2007), or cause air passengers to spill their drinks.  These movements, which depend on the characteristics of the vehicle as well as the wind shear, are typically referred to as 'turbulence'. Turbulence more generally, however, is the spatio-temporal variation of the wind, and can also manifest itself as fluctuations in a time series of wind measured at a fixed point as the turbulent wind field is advected past a meteorological station.  These fluctuations are of interest in that they are intimately associated with the transport of matter, energy and momentum at the macroscale through an atmosphere via so-called eddy diffusivity. Turbulent fluctuations are often also the critical factor in exceeding a transport threshold where mean winds are not enough (e.g. the saltation speed for sand grains – Burr et al., 2015; Lorenz et al., 2014;  Tokano et al, 2010).  The fluctuations also represent the process by which mechanical energy is ultimately dissipated by the turbulent cascade into viscous dissipation.

Here I explore what the Huygens descent measurements indicate about these topics, most particularly through the wind measurements made by the Doppler Wind Experiment (Bird et al., 2005; Folkner et al., 2006).  First, however, it is pertinent to recall what kind of wind shear environment was anticipated.

2. Pre-Cassini Expectations of Wind Shear

Because the gust environment could excite motions of the Huygens probe under its parachute, thereby causing an interruption of its radio link to Cassini by mispointing its antennas, an evaluation of wind shear was made during the Huygens probe development  (Strobel and Sicardy, 1997).   Having to devise a specification on some aspect of an planetary environment for a vehicle intended to make the very first in-situ measurements of that environment is one of the great ironical challenges of planetary exploration. The evaluation by Strobel and Sicardy (1997) identified physical constraints on wind shear (specifically, the vertical gradient of horizontal winds), suggesting that a strong upper limit on shear determined by the critical Richardson number was 2N, where N is the Brunt-Vaisala (buoyancy) frequency, determined by the vertical stability of the atmosphere. Although not intuitively obvious, it is seen that the buoyancy frequency units of inverse seconds are the same as those of wind shear, namely meters per second per meter. The buoyancy frequency profile was estimated from the temperature structure derived for Titan's low-latitude atmosphere from the Voyager 1 radio occultation experiment. The peak shear (2N) anticipated was 16 m/s/km at an altitude of 60km. As this paper shows later, this estimate, devised from very limited information, proved to be a generally appropriate one.

In flowing down this expectation of the maximum possible shear to a parachute engineering specification (table 1, from Aerospatiale, 1997), the profile was de-rated somewhat by 'engineering judgement'. This was presumably because it was hard to meet underlying requirement that "the probe shall limit the pendular motion during the descent to an amplitude of $10°$" in that in steady state descent at speed W, if the wind shear is $\gamma$, then the probe-parachute system will trim at an angle $\Theta$ given by $\tan(\Theta) = W\gamma/g$. Since the terminal velocity W at 60km is 25 m/s, adopting the upper limit of 2N would give a trim of $16°$ - thus even with a parachute system that was perfectly damped, the allowed limit would be violated. The specification was therefore rationalized by cutting off the peak and reducing the profile overall. The specification was considered a '95%' limit (without documented statistical justification, but with the intent that occasional violations could be tolerated).

Table 1. Huygens Wind Shear Specification

| Altitude (km) | 95% Wind Gradient m/s/km |
|---|---|
| <10 | 3.1 |
| 10 | 3.1 |
| 20 | 4.6 |
| 30 | 5.6 |
| 40 | 6.3 |
| 50 | 7.1 |
| 60 | 9.2 |
| 70 | 9.1 |
| 80 | 7.3 |
| 90 | 6.2 |
| 100 | 5.6 |
| 110 | 5.2 |
| 120 | 5.1 |
| 130 | 5.0 |
| 140 | 4.8 |
| 150 | 4.7 |
| 160 | 4.5 |

3. Wind Profile in the Free Atmosphere

The principal source of wind information on Titan is from the Doppler tracking of the Huygens probe (Bird et al., 2005; Folkner et al. 2006). The range-rate (i.e. velocity along the line-of-sight to the Earth) was measured using a precision frequency reference on the radio link. With knowledge of the viewing geometry and an estimate of the vertical descent speed of the probe (which had a significant projection on the line-of-sight) and with the assumption that meridional winds were zero, this range rate could be rescaled into a zonal (E-W) probe velocity, generally close to the ambient zonal wind velocity. The resultant dataset (HP-SSA-DWE-2-3-DESCENT-V1.0) is archived on the Atmospheres Node of the Planetary Data System.

In fact, the meridional velocity is not quite zero, and some initial estimates were made by computing the probe location relative to features seen on the ground by the Descent Imager and Spectral Radiometer (DISR) on Huygens during the latter part of descent (Tomasko et al., 2005). Recently, an improved evaluation of meridional winds has been made using the entire descent image dataset (Karkoschka, 2016). While these data are useful to define the overall wind profile (and indicate a shear of 1 m/s/km in the lowest 2km of the atmosphere and very little shear above), they are too sparse to meaningfully evaluate fluctuations. The meridional wind speed is small enough that the assumption of zero i reduction of the range rate measurement to a zonal wind speed is essentially unaffected.

The zonal wind profile is plotted on figure 1, and several broadly linear regions are identified, with large-scale (several km in vertical extent) shears of up to 5 m/s/km. This observed shear profile is plotted as grey bars in figure 2, which also shows the (doubled) buoyancy frequency from Strobel and Sicardy (1997) and the adopted specification from table 1. It is seen that the specification was not violated, although there was essentially zero margin in the 110-120km altitude range. The original recommendation of a robust upper limit of 2N would have had approximately 50% margin.

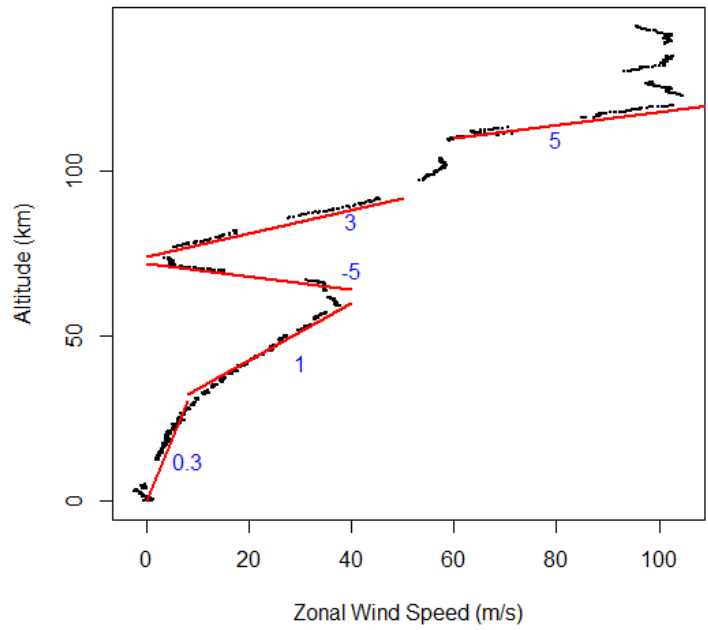

Figure 1. The Huygens Doppler Wind profile (dots). Approximately linear regions of wind shear are noted, with the gradient (in m/s/km) noted.

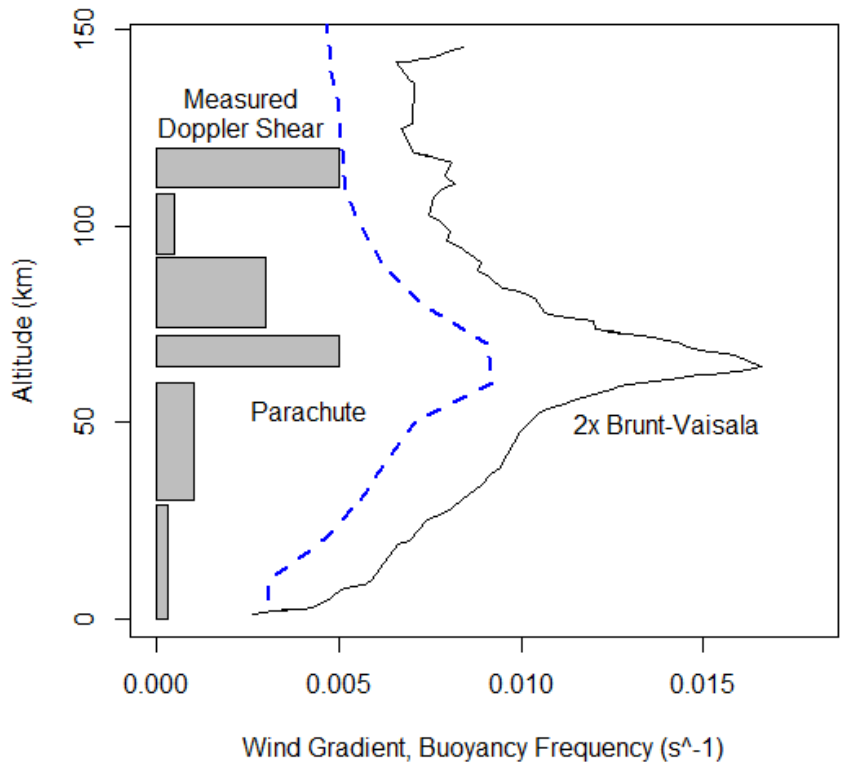

Figure 2. The upper limit on expected wind shear defined in Strobel and Sicardy (1997, solid curve) as twice the Brunt-Vaisala frequency, compared with the adopted 95% Huygens wind shear specification (blue dashed curve, table 1) and the observed shear from the linear segments identified in figure 1 (grey bars). It is seen that in general the wind shear did not quite exceed the specification. It is of interest that one region of strongest observed shear (60-80km) was where the shear was anticipated to be strongest.

To put these shears in context, although turbulence in passenger aviation is usually defined in terms of vehicle motion (and a small, slow-flying aircraft will typically respond more violently to a given wind shear than a heavy, fast-flying one), Houboldt's review paper on the topic (Houboldt, 1973) suggests that 'Light' turbulence is typically associated with a shear of 2.5-10 m/s/km. Thus some parts of the

Huygens descent met this classification, but moderate or severe (>17.5 m/s/km) shear levels were not encountered.

4. Near-Surface Atmosphere Profile

The atmosphere near the ground is considered separately from the free atmosphere : the stability of the Planetary Boundary Layer (PBL) makes its behaviour quite distinct. The PBL is indicated in the potential temperature profile (temperature with an adiabatic correction for altitude ) which shows a constant region up to 300m altitude (figure 3). Although initially interpreted as defining 'the planetary boundary layer' (Tokano et al., 2006) this seems in fact merely to be the top of the growing diurnal boundary layer – the Huygens descent was at a local solar time of about 9am.

Lorenz et al. (2010) noted that breaks in slope higher in the profile may in fact indicate the full depth of the layer relevant for near-surface atmospheric dynamics, and in particular breaks at 2 & 3.5km (C and D) in figure may be responsible for controlling the remarkably regular spacing of Titan's dunes. Dunes form small and, given enough time in a regular wind pattern and enough sediment supply, progressively grow. Thus an instantaneous view may simply represent the history of the dune field since climate cycles established a wind regime, but ultimately dune growth is capped by the boundary layer which seems to control the size and spacing of the largest dunes on Earth. More recently Charnay and Lebonnois (2012) have simulated boundary layer growth in a global circulation model and find that the diurnal boundary layer may reach about 1km in thickness (possibly inflection B in figure) whereas a seasonal layer grows to a couple of times thicker than that. These structures have not previously been related to the Doppler wind profile.

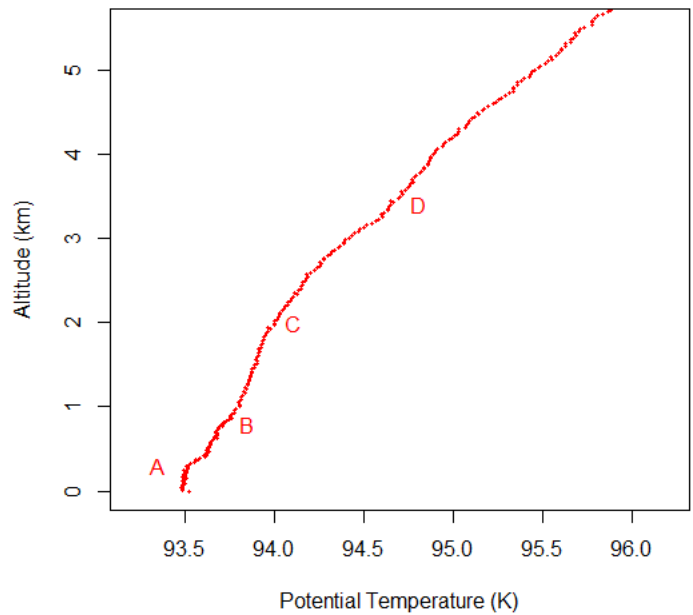

Figure 3. The potential temperature profile (Tokano et al., 2006; Lorenz et al., 2010) shows several features. A is interpreted as the top of the growing morning boundary layer at 300m ; GCM simulations attribute B to the vestige of the previous day's fully-developed diurnal boundary layer (Charnay and Lebonnois, 2012) ; C or D may be a seasonal boundary layer (Charnay and Lebonnois, 2012) : D appears to control the size and spacing of Titan's dunes.

Inspection of the Doppler wind profile (figure 4) indicates two changes in shear sign at about 0.3 and 3 km, which appear coincident with the A and D discontinuities in the potential temperature profile. There does not appear to be a major feature at 1km, although this may not be especially significant since the wind speed was approximately zero at this altitude. The km-scale shears are generally smaller than in the free atmosphere : 1 and 2.5 m/s/km. There is as one might expect a shear near the surface due to surface friction, possibly as large as 10 m/s/km. However, the wind speed in the lowest km was in any case only 1m/s or less and the measurement uncertainty (also included in the PDS dataset) becomes significant – discarding a single data point would bring the upper limit of shear down to only 5 m/s/km. There are point-to-point variations throughout the profile that often exceed the error bars, indicating some level of small-scale turbulence.

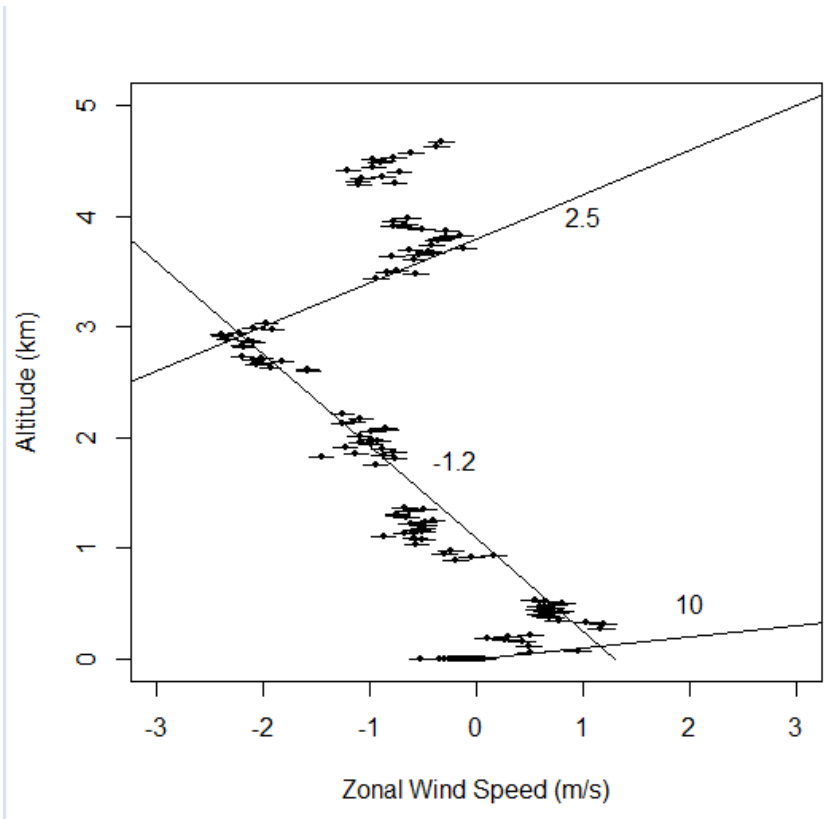

Figure 4. Doppler measurements (with estimated error bars) in the lowest 5km of Titan's atmosphere. Gaps occur due to the need to nod the radio telescope periodically to celestial reference sources. The wind is rather small (<2 m/s) and reverses direction a couple of times, but with gradients <2 m/s/km, except possibly just at the surface. Here (indicated by only one or two data points) the gradient may be 10 m/s/km. The two changes in shear sign are interestingly coincident with the main breaks in potential temperature structure A and D.

5. Turbulence Measurements and Dissipation Rate

Examining these small-scale fluctuations, we can interpret the Doppler Wind profile near the surface in an alternative way. It is instructive to ignore the vertical descent of the vehicle, and merely interpret the measurements as a wind time series. Taking differences between successive measurements (which for the lowest part of descent, where the Doppler measurements were made at the Parkes radio telescope in Australia, correspond to 3-second integrations) correspond to yields an approximately Gaussian distribution (figure 5). Of 76 differences, 6 are above 0.4 m/s, and one is above 0.6 m/s, broadly consistent given the poor statistics with these being considered '2-sigma' (95%) and '3-sigma' values.

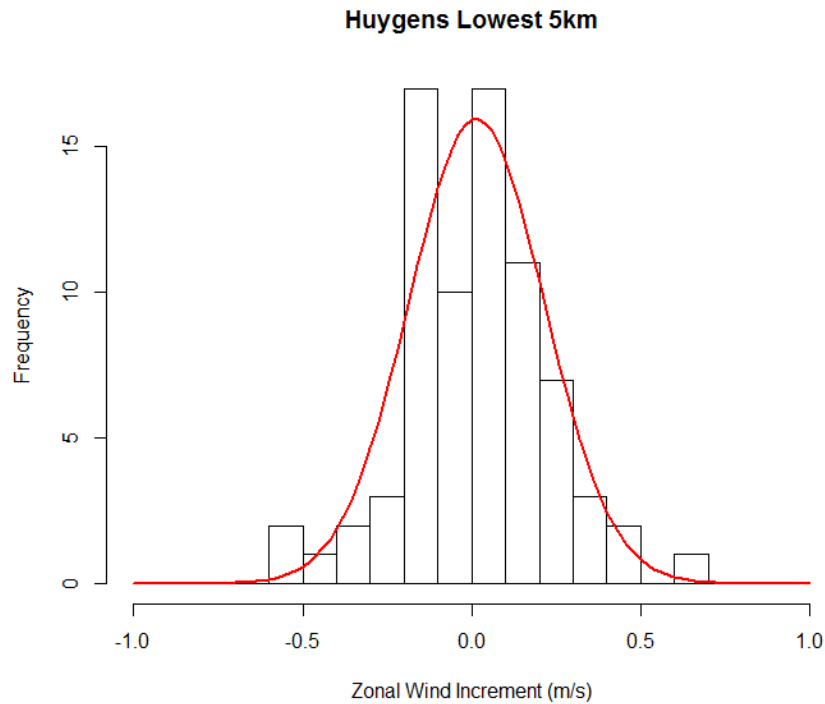

Figure 5. Differences between successive Doppler wind measurements (3 seconds apart, histogram) are approximately Gaussian (red line) with a standard deviation of 0.21 m/s.

There is independent information suggesting a characteristic fluctuation of this order in the lower atmosphere. Tilt sensors on the probe (Lorenz et al., 2007) were sampled at 1-second intervals, and while the data were influenced by short-period dynamical effects in the earlier, faster part of descent, near the surface where the descent speed was only ~5 m/s, tilts and angular rates were low (Lorenz, 2009). Examining the data (archive product SSP_TIL_123456_0_R_ATMOS.TAB) shows the tilts to be Gaussian in the lowest 4km with a standard deviation of 2.5 degrees (supported by Karkoschka's (2016) analysis of DISR images indicating tilts of ~2° in the lowest part of descent.) Lorenz (2007) showed that for small tilts $\alpha$, the turbulent wind fluctuations $\sigma_u$ can be estimated as $\sigma_u \sim \alpha W/2$, and thus for a descent speed W=5 m/s and a~0.04 radians, $\sigma_u$~0.1 m/s.

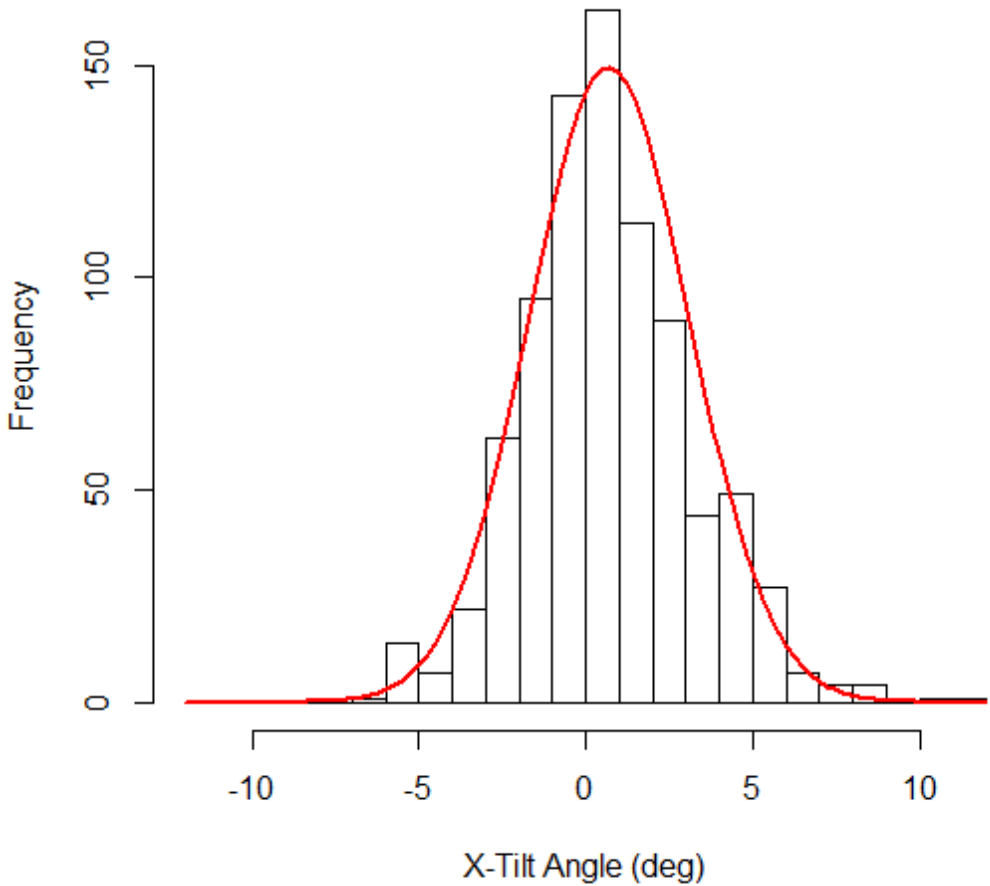

Figure 6. Huygens tilt sensor readings in the lowest 4 km are Gaussian with standard deviation of 2.5°, which can be interpreted as a wind fluctuation of ~0.1 m/s (see text)

It is of reassuring interest that the characteristic fluctuation is comparable with the convective velocity scale (Deardorff 1970) which can be written $w* = [(g/T)z(F/\rho c_p)]^{0.33}$, where g is acceleration due to gravity, T is the absolute temperature, z is the height of the convecting layer, F is the sensible heat flux and $c_p$ is the specific heat. The terms in parentheses represent a buoyancy parameter and the kinematic heat flux respectively. Adopting, conservatively, z=3km, and F=0.7 W/m² (e.g. Williams et al., 2012) we find w*=0.15 m/s.

The formulation above is derived from wind measurements at a few km altitude, where the actual wind speed was around 1 m/s thus the simple 'turbulence intensity' metric $\sigma/u$ in this instance is 10-20%. This may be compared with the models for low altitude turbulence in terrestrial aviation. For example, aircraft handling standard MIL-STD-1797 defines the value $\sigma_w/w$ (vertical wind) at an altitude of 20 ft (~7m)) to be 0.1, and indicates horizontal wind turbulence to be double this value. A more extensive compilation is given by Lappe (1966) who gives expressions of the form $\sigma_w$ = A+Bu for a range of different terrains and stability conditions derived from low-altitude B-66 flights, with A=0 to 3 m/s and B=0.0 to 0.2.

A handful of turbulence measurements have been reported on other planetary bodies. On Mars, Holstein-Rathlou et al. (2010) reported turbulence levels $\sigma_u/u$ at the Phoenix lander to be ~0.1 at 9 and 17 hrs local solar time, rising to ~0.35 in the middle of the day. Golitsyn (1978) reports the mean and root-mean-square fluctuations in anemometer readings to be 0.4 and 0.1 m/s for Venera-9 and 0.9 and 0.15 m/s for Venera-10 : these correspond to $\sigma_u/u$ of 0.25 and 0.17 respectively. Thus the Huygens measurements seem broadly typical.

Golitsyn (1978) derived a turbulent dissipation rate (at 1.3m altitude) of 1.3 and 4.3 cm$^2$/sec$^3$ for these two sites on Venus. Kerzhanovich et al. (1980) performed Doppler tracking of Venera 11 and 12, and derived dissipation rates in the lower Venusian atmosphere of 3-9 cm$^2$/sec$^3$. If we adopt a length scale L of ~500m for the lowest-altitude set of Doppler measurements on Huygens, and $\sigma_u$ ~ 0.2m/s, then the turbulent dissipation rate $\varepsilon=(\sigma^3_u/L)$ is ~16 cm$^2$/sec$^3$. For comparison, on Earth with a wind speed of 5 m/s, Priestley (1959) suggests $\varepsilon=10^2$-$10^4$ cm$^2$/sec$^3$ at 1m height (see also Taylor, 1952), and 1-10 for 500m. Thus in broad terms, Titan appears to have more near-surface dissipation at a given height than Venus, but less than Earth, although all three worlds are broadly comparable.

6. Simulated Turbulent Wind

Although the variations in wind speed have been described above in terms of a standard deviation, it should not be assumed that wind can be modeled (e.g. to simulate the dynamics of exploration vehicles at Titan, or to generate synthetic measurement datasets) by a purely random series. There is some persistence in the signal, which can be succinctly captured in a simple order-one autoregressive model AR(1).  Specifically, the discrete time series  U(t) with t=1,2,3 seconds,.. is generated as

$$U(t)=pU(t-1)+N(0,s) \qquad (1)$$

where p is the autoregression coefficient, and N(0,$\sigma$) is a Gaussian-distributed random variable with zero mean and standard deviation $\sigma$.  In the case of p=0, the signal is uncorrelated white noise, while p=1 defines the Weiner process, resulting in classical Brownian motion.  Inspection of the Huygens Doppler data using the  "arima" function in the statistical analysis package "R" suggests here q=0.96, implying a near-Brownian characteristic.

Turbulence models used in terrestrial aeronautics (e.g. Houboldt, 1973) are essentially similar in form, in that time series with the relevant statistical properties can be generated by the linear filtering of a white noise signal. Fichtl (1978) provides discrete equations identical in form to eq.1 in his turbulence model for the Space Shuttle orbiter.

A synthetic time series of turbulent fluctuations is generated according to the model above, with $\sigma$ =0.06 m/s   (corresponding to 1-second increments, since the deviation of 0.2 m/s earlier corresponded to 3-second integrations), is transformed into an altitude profile roughly as h(t)=0.0047t  (the terminal descent of Huygens actually declined from 4.9 m/s at 4km to 4.5 m/s at the surface) and  is superposed on a linear gradient, U(h)=0.75-0.85h  to yield the profile in figure 7.   It is seen that the variations are rather plausibly described by this simple model.

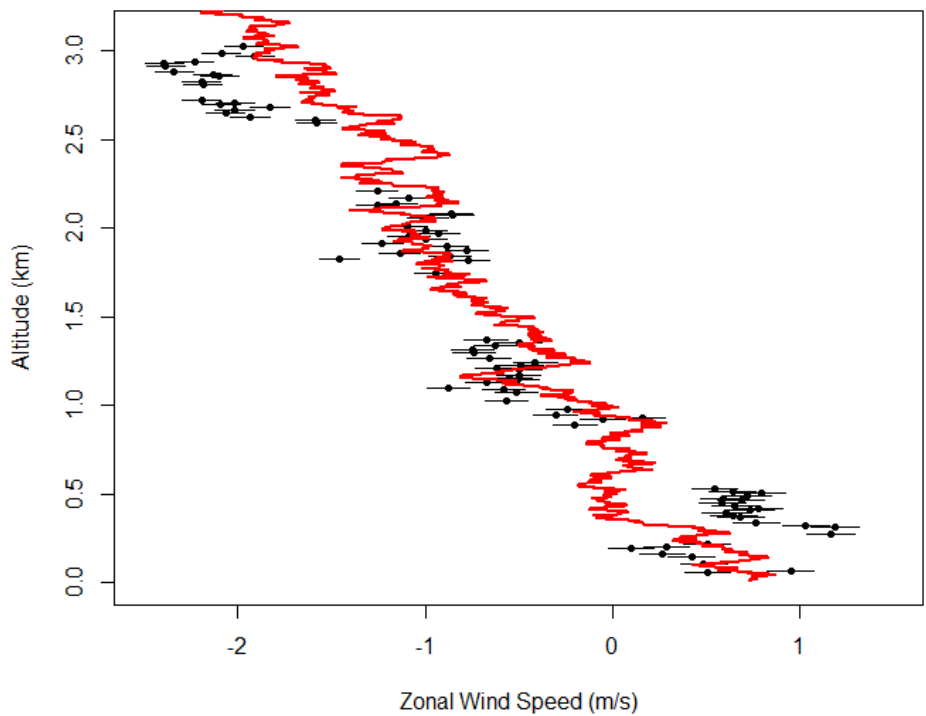

Figure 7. The Huygens Doppler measurements in the lowest 3km of the atmosphere (points) compared with a linear altitude trend with a superposed AR(1) time series of simulated turbulent fluctuations as described in the text. The model variations appear consistent with those observed.

7. Conclusions

Huygens flight data have been reviewed to assess wind shear and turbulence. Shear was within pre-flight limits (although only just), and reached levels typical for 'light turbulence' in aviation. Near-surface changes in wind shear at ~300m and 3km appear to coincide with kinks in the potential temperature profile associated with planetary boundary layer features. Turbulent fluctuations in the lowest few km of the atmosphere of the order of 0.1-0.2 m/s over timescales of 1-3 s are observed, and a simple AR(1) model can simulate the observed turbulent structure. These observations may be useful in future Titan exploration.

Acknowledgement

This work was supported in part by NASA Grant "Cassini Science Support" NNX13AH14G.

References

Collet, C. 1997. Huygens Probe User Manual Description, HUY.AS/c.100.OP.0201 Issue 4 Rev/B, Aerospatiale, Cannes, France, 15 September 1997

Bird, M.K., Allison, M., Asmar, S.W., Atkinson, D.H., Avruch, I.M., Dutta-Roy, R., Dzierma, Y., Edenhofer, P., Folkner, W.M., Gurvits, L.I. and Johnston, D.V., 2005. The vertical profile of winds on Titan. Nature, 438, 800-802

Burr, D.M., Bridges, N.T., Marshall, J.R., Smith, J.K., White, B.R. and Emery, J.P., 2015. Higher-than-predicted saltation threshold wind speeds on Titan. Nature, 517, 60-63.

Charnay, B. and Lebonnois, S., 2012. Two boundary layers in Titan's lower troposphere inferred from a climate model. Nature Geoscience, 5, 106-109.

Deardorff, J. W. 1970. Convective velocity and temperature scales for the unstable planetary boundary layer and for Rayleigh convection. J. Atmos Sci. 27, 1211–1213

Dzierma, Y., Bird, M.K., Dutta-Roy, R., Pérez-Ayúcar, M., Plettemeier, D. and Edenhofer, P., 2007. Huygens Probe descent dynamics inferred from Channel B signal level measurements. Planetary and Space Science, 55, 1886-1895.

Fichtl, G.H., 1977. A technique for simulating turbulence for aerospace vehicle flight simulation studies. NASA Technical Memorandum, NASA TM 78141, Marshall Space Flight Center, Alabama, November 1977

Folkner, W.M., Asmar, S.W., Border, J.S., Franklin, G.W., Finley, S.G., Gorelik, J., Johnston, D.V., Kerzhanovich, V.V., Lowe, S.T., Preston, R.A. and Bird, M.K., 2006. Winds on Titan from ground-based tracking of the Huygens probe. Journal of Geophysical Research: Planets, 111(E7) doi:10.1029/2005JE002649

Golitsyn, G.S., 1978. Estimates of the turbulent state of the atmosphere near the surface of Venus from the data of Venera 9 and Venera 10. Cosmic Research, 16, 125-127

Holstein-Rathlou, C., H.P. Gunnlaugsson and 19 co-authors (2010) Winds at the Phoenix landing site, J. Geophys. Res., 115, E00E18, doi: 10.1029/2009JE003411

Houboldt, J. C. 1973. Atmospheric Turbulence, AIAA Journal, 11, 421-43

Karkoschka, E., 2016. Titan's meridional wind profile and Huygens' orientation and swing inferred from the geometry of DISR imaging. Icarus, 270, 326-338

Kerzhanovich, V.V., Mararov, Y.F., Marov, M.Y., Rozhdestvenskiy, M.K. and Sorokin, V.P., 1980. Venera 11 and Venera 12: Preliminary evaluations of wind velocity and turbulence in the atmosphere of Venus. The Moon and the Planets, 23, 261-270

Lappe, U. O., 1966. Low-Altitude Turbulence Model for Estimating Gust Loads on Aircraft, Journal of Aircraft, 3, 41-47

Lingard, J.S., Underwood, J.C., Darley, M.G., Merrifield, J., Caldwell, J., Paris, S., Bányai, T. and Molina, R., Huygens Flight Performance Analyses. In 8th European Symposium on Aerothermodynamics for Space Vehicles. Lisbon, Portugal 2-6th March 2015 (www.congrexprojects.com/15A01, downloaded 26/7/2016)

Lorenz, R. D. Attitude and Angular Rates of Planetary Probes during Atmospheric Descent, Planetary and Space Science, 58, 838-846, 2010

Lorenz, R.D., Claudin, P., Andreotti, B., Radebaugh, J. and Tokano, T., 2010. A 3km atmospheric boundary layer on Titan indicated by dune spacing and Huygens data. Icarus, 205, 719-721.

Lorenz, R.D., 2014. Physics of saltation and sand transport on Titan: A brief review. Icarus, 230, 162-167

Lorenz, R. D., J. Zarnecki, M. C Towner, M. Leese, A. Ball, B. Hathi, A. Hagermann, N. Ghafoor, 2007. Descent Motions of the Huygens Probe as Measured by the Surface Science Package (SSP) : Turbulent Evidence for A Cloud Layer, Planetary and Space Science, 55, 1936-1948

Lorenz, R. D., P. Claudin, J. Radebaugh, T. Tokano and B. Andreotti, 2010. A 3km boundary layer on Titan indicated by Dune Spacing and Huygens Data, Icarus , 205, 719–721

Priestley, C.H.B., 1959. Turbulent transfer in the lower atmosphere, University of Chicago Press, Chicago, 60-62

Strobel, D. F. and B. Sicardy, 1997. Gravity Wave and Wind Shear Models, 299-311 in Wilson, A. (Ed.), Huygens, Science, Payload and Mission, ESA SP-1177. European Space Agency, Noordwijk, The Netherlands

Taylor, R.J., 1952. The dissipation of kinetic energy in the lowest layers of the atmosphere. Quarterly Journal of the Royal Meteorological Society, 78, 179-185

Tokano, T., 2010. Relevance of fast westerlies at equinox for the eastward elongation of Titan's dunes. Aeolian Research, 2, 113-127.

Tokano, T., Ferri, F., Colombatti, G., Mäkinen, T. and Fulchignoni, M., 2006. Titan's planetary boundary layer structure at the Huygens landing site. Journal of Geophysical Research: Planets, 111(E8). doi:10.1029/2006JE002704

Tomasko, M.G., Archinal, B., Becker, T., Bezard, B., Bushroe, M., Combes, M., Cook, D., Coustenis,A., de Bergh, C., Dafoe, L.E., Doose, L., Douté, S., Eibl, A., Engel, S., Gliem, F., Grieger, B., Holso, K., Howington Kraus, E., Karkoschka, E., Keller, H.U., Kirk, R., Kramm, R., Kuyppers, M., Lanagan, P., Lellouch, E., Lemmon, M., Lunine, J., McFarlane, E., Moores, J., Prout, G.M., Rizk, B., Rosiek, M., Rueffer, P., Schroeder, S.E., Schmitt, B., See, C., Smith, P., Soderblom, L., Thomas, N., West, R., 2005. Rain, winds and haze during the Huygens probe's descent to Titan's surface. Nature 438, 765–778

Williams, K.E., McKay, C.P. and Persson, F., 2012. The surface energy balance at the Huygens landing site and the moist surface conditions on Titan. Planetary and Space Science, 60, 376-385